\documentstyle[12pt]{article}
\begin{document}
\topmargin 0pt
\oddsidemargin 5mm
\setcounter{page}{1}
Preprint YERPHI-1473(10)-96\\
\vspace{2cm}
\begin{center}
{SELFACCELERATION OF ELECTRONS IN ONE-DIMENSIONAL}\\
{BUNCHES,MOVING IN COLD PLASMA}\\
{\bf A.Ts. Amatuni}\\
{\it Yerevan Physics Institute,}\\
{\it Alikhanian brothers st.2,Yerevan 375036,}\\
{\it Republic of Armenia}
\end{center}
\vspace{5mm}
\centerline{Abstract}
{Nonlinear dynamics of the one-dimensional ultrarelativistic bunch of
electrons,moving in cold plasma,is considered in multiple scales perturbative
approach.A square root of the inverse Lorentz factor of the bunch electrons
is taken as a small parameter.Bunch electrons momenta is changed in the
first approximation.In the underdense plasma and for the model example of the
combined bunch the selfacceleration of the bunch electrons can be remarkable.

\section{Introduction}

Nonlinear wake waves exitation in overdense plasma by relativistic electron
(positron) one-dimensional bunches,when it is possible to obtain an exact
analytical solution,was considered in \cite{A}-\cite{G}.The bunch 
assumed as a given one (rigid bunch approximation) in the most 
of these work.

The buck influence of the exited plasma wake on electron bunch was 
considered numericaly in \cite{G},\cite{H}.Some attempts of the analytical
treatment of the problem have been performed in \cite{H}-\cite{L}.

In the present work the problem of the plasma back nonlinear influence
on the driving one-dimensional electron bunch is treated by the method of
multiple scales \cite{M}.The bunch is ultrarelativistic  and the square
root of the inverse Lorentz factor of the bunch $\epsilon \equiv 
\gamma_0^{-1/2} \ll 1$ is taken as a small parameter.

It is assumed that the bunch-plasma interaction takes place by two stages.
First one is the formation of the stationary plasma wake field,generated 
by the rigid bunch and the second stage is the influence on the momenta of
the bunch electrons and wake field itself by this field.This assumption
is valid,when $\gamma_0 \gg 1$.Indeed,the time $\tau_s$ needed for the 
formation of the stationary wake in a plasma with density $n_0$,generated
by the electron bunch with the density $n_b,n_b/n_0<1/2-\triangle,
\frac{1}{8}\gamma_0^{-2}\ll \triangle<1/2$,is $\tau_s \sim \omega_p^{-1},
\omega_p^2=4\pi e^2n_0/m$.The time required for the change of the bunch
electrons momenta $p_0$ is $\tau_p \sim p_0/eE_0$ where $E_0$ is the electric
field inside the bunch.In the overdense $(\frac{n_b}{n_0}<1/2)$ plasma
$E_0 \leq \frac{mc\omega_p}{e}$;in the underdens ($\frac{1}{2}n_0 \ll n_b$)
plasma $E_0 \leq \frac{mc\omega_p}{e}\gamma_0^{1/2}$,(\cite{E};see for 
details Section 6 below).Hence in the overdense plasma $\tau_s/\tau_p \leq
\gamma_0^{-1} \ll 1$,and in the underdense plasma $\tau_s/\tau_p \leq
\gamma_0^{-1/2} \ll 1$.(For some special values of $\frac{n_b}{n_0} \simeq
1/2$ these conditions can change the form or even violate).As in \cite{E}
consider the flat electron bunch with the infinite 
transverse dimensions,longitudinal length $d$ and initial homogenious
charge density $n_b$,moving in the lab system with the initial velocity
$v_0$ through neutral cold plasma with the immobile ions.

In the work \cite{N} it was shown that when the beam is traversing the
semiinfinite plasma after a few plasma wave lengths transient effects
dissipate and stationary wake field regime is established
in coinsideness with the abovementioned estimate.
In what follows
this moment is taken as an initial one,$t=0$,and the future development
for $t>0$ of the bunch -plasma system is considered.

\section{Formulation of the Problem}
The considered electron bunch -cold plasma system is described by the 
hydrodynamic equations of the motion,continuity equations for charge 
densities and currents for the bunch and plasma electrons,and Maxwell
equation (Coulomb law) for the electric field.

Dimensionless variables and arguments are introduced:
\begin{eqnarray}
\label{A}
t'=\omega_pt,z'=k_pz,\omega_p^2=\frac{4\pi e^2n}{m},k_p=\omega_p/c,\\ 
\nonumber
E=(4\pi nmc^2)^{1/2}E'=\frac{mc\omega_p}{e}E',{n'}_e=\frac{n_e}{n},
{n'}_b=\frac{n_b}{n} \\ \nonumber
\rho_e=\frac{p_e}{p_b},\rho_b=\frac{p_b}{mc\gamma_0},\beta_b=
\frac{v_b}{c},\beta_e=\frac{v_e}{c}, \\ \nonumber
\gamma_0=(1-\beta_0^2)^{-1/2},
\end{eqnarray}
where $p_e,v_e,n_e;p_b,v_b,n_b$ are the momenta,velocity and density of the
plasma and bunch electrons subsequently,$n$ is a normalizing 
constant,which is suitable to choose $n=n_0$ for overdense 
plasma and $n=n_b$ for underdense plasma cases,$\beta_0=v_0/c$
is the initial velocity of the bunch electrons in the 
laboratory system.

The equations,which describe the considered problem are:
\begin{equation}
\label{B}
\frac{\partial{p_e}}{\partial{t}}+\beta_e\frac{\partial{p_e}}
{\partial{z}}=-E
\end{equation}
\begin{equation}
\label{C}
\frac{\partial{p_b}}{\partial{t}}+\beta_b\frac{\partial{p_b}}
{\partial{z}}=-\frac{1}{\gamma_0}E
\end{equation}
\begin{equation}
\label{D}
\frac{\partial{n_e}}{\partial{t}}+\frac{\partial}
{\partial{z}}(\beta_e n_e)=0
\end{equation}
\begin{equation}
\label{E}
\frac{\partial{n_b}}{\partial{t}}+\frac{\partial}
{\partial{z}}(\beta_b n_b)=0
\end{equation}
\begin{equation}
\label{F}
\frac{\partial{E}}{\partial{z}}=n_0-n_e-n_b
\end{equation}
Eqs. (\ref{B}-\ref{F}) are written in dimensionless variables (\ref{A})
and prime is omitted.Considering the ultrarelativistic bunch,introduce
a small parameter $\epsilon^2=1/\gamma_0$,and all variables,entering in
(\ref{B}-\ref{F}),let be a functions of the fast $\tilde{z}=z-\beta_0 t$ and
slow $\tau=\epsilon t,\zeta=\epsilon z$ arguments.According to the multiple
scale method \cite{M},all variables,entered in (\ref{B}-\ref{F}),are 
developed in the following serieses:
\begin{equation}
\label{G}
\rho_e=\rho_{e0}(\tilde{z})+\epsilon\rho_{e1}(\tilde{z},\zeta,\tau)+
\epsilon^2\rho_{e2}(\tilde{z},\zeta,\tau)+\cdots
\end{equation}
\begin{equation}
\label{H}
\beta_e=\beta_{e0}(\tilde{z})+\epsilon\beta_{e1}(\tilde{z},\zeta,\tau)+
\epsilon^2\beta_{e2}(\tilde{z},\zeta,\tau)+\cdots
\end{equation}
\begin{equation}
\label{I}
\rho_b=\rho_{b0}+\epsilon\rho_{b1}(\tilde{z},\zeta,\tau)+
\epsilon^2\rho_{b2}(\tilde{z},\zeta,\tau)+\cdots
\end{equation}
\begin{equation}
\label{J}
\beta_b=\beta_{b0}+\epsilon\beta_{b1}(\tilde{z},\zeta,\tau)+
\epsilon^2\beta_{b2}(\tilde{z},\zeta,\tau)+\cdots
\end{equation}
\begin{equation}
\label{K}
n_e=n_{e0}(\tilde{z})+\epsilon n_{e1}(\tilde{z},\zeta,\tau)+
\epsilon^2 n_{e2}(\tilde{z},\zeta,\tau)+\cdots
\end{equation}
\begin{equation}
\label{L}
n_b=n_{b0}+\epsilon n_{b1}(\tilde{z},\zeta,\tau)+
\epsilon^2 n_{b2}(\tilde{z},\zeta,\tau)+\cdots
\end{equation}
\begin{equation}
\label{M}
E=E_0(\tilde{z})+\epsilon E_1(\tilde{z},\zeta,\tau)+
\epsilon^2 E_2(\tilde{z},\zeta,\tau)+\cdots
\end{equation}

In (\ref{G}-\ref{M}) $\rho_{b0},\beta_0,n_{b0}$ are the constants
connected with the initialy rigid electron bunch and differs from zero,
when $0 \leq \tilde{z} = z-\beta_0 t \leq d$. The functions 
$\rho_{e0}(\tilde{z}),\beta_{e0}(\tilde{z}),n_{e0}(\tilde{z}),
E_0(\tilde{z})$ are the solutions of the steady state (stationary)
problem,and are obtained in \cite{A},\cite{C},\cite{E}.Derivatives in
(\ref{B}-\ref{F}),according to the multiple scales method are given by
\begin{eqnarray}
\label{N}
\frac{\partial}{\partial t}=-v_0\frac{\partial}{\partial \tilde{z}}+
\epsilon \frac{\partial}{\partial \tau} \\ \nonumber
\frac{\partial}{\partial z}=\frac{\partial}{\partial \tilde{z}}+
\epsilon \frac{\partial}{\partial \zeta}
\end{eqnarray}

Decompositions (\ref{G}-\ref{N}) correspond to the above mentioned main 
assumption,which,as it was seen,is valid for the ultrarelativistic bunch.

The following steps are evident-- decomposition of the eqs. (\ref{B}-
\ref{F}),using (\ref{G}-\ref{N}) provides the sequence of the guasilinear
equations,describing the steady state regime as a zero approximation and
in the next approximations-the back influence of the generated electric field
on initialy rigid electron bunch and on wake wave itself. 

\section{Dynamics of the Bunch Electrons}

It is necessary for the following to know the relations between $\rho_{bi}$
and $\beta_{bi}$ (\ref{I}-\ref{J}).From the definition (\ref{A}) it follows
that:
\begin{equation}
\label{O}
\beta_b=\left(1+\frac{{\epsilon}^4}{\rho_b^2}\right)^{-1/2}=
\beta_0+{\epsilon}^5\beta_{b5}+{\epsilon}^6\beta_{b6}+{\epsilon}^7\beta_{b7}+
\cdots
\end{equation}

It is essential to notice that in (\ref{O}) the terms proportional to
${\epsilon}^1,\dots,{\epsilon}^4$ are absent.Using (\ref{I}) and (\ref{O})
we have
\begin{equation}
\label{P}
\beta_{b5}=\frac{\rho_{b1}}{\beta_0^3} \qquad (\rho_{b0}=\beta_0)
\end{equation}
\begin{equation}
\label{Q}
\beta_{b6}=-\frac{1}{2\beta_0^3}(3\rho_{b1}^2-2\rho_{b2}\rho_{b0})
\end{equation}
\begin{equation}
\label{R}
\beta_{b7}=-\frac{1}{\beta_0^3}(3\rho_{b1}\rho_{b2}+2\rho_{b1}^3\rho_{b0}-
 \rho_{b3}\rho_{b0})
\end{equation}

Zero $(\sim \epsilon^0)$ and first order $(\sim \epsilon)$ equations 
(\ref{C}) for bunch electron momenta identically satisfied.The second
order equations $(\sim \epsilon^2)$ is
\begin{equation}
\label{S}
\frac{\partial \rho_{b1}}{\partial \tau}+\beta_0\frac{\partial \rho_{b1}}
{\partial \zeta}=-E_0(\tilde{z}),
\end{equation}
where $E_0(\tilde{z})$ is the solution of the steady state problem 
presented in \cite{E}.

The characteristics of the eq. (\ref{S}) are 

$$\zeta-\beta_0\tau=c_1;\rho_{b1}+E_0(\tilde{z})\tau=c_2;$$
Boundary condition is $\rho_{b1}=0$,when $\tilde{z}=d;$initial 
condition
is $\rho_{b1}=0$,when $\tau=0;\rho_{b1} \neq 0,$when $\epsilon d_0 <
\zeta-\beta_0\tau < \epsilon d,$where $d_0$ is the position of the end
of the bunch for $t>0;d_0=0$,when $t=0$;the solution of the eq. (\ref{S})
is
\begin{equation}
\label{T}
\rho_{b1}=-E_0(\tilde{z})\tau[\theta(\zeta-\beta_0\tau-\epsilon d_0)-
\theta(\zeta-\beta_0\tau-\epsilon d)]
\end{equation}
In the usual units eq. (\ref{T}) reads
\begin{equation}
\label{U}
p_{b1}=-eE_0(\tilde{z})t\gamma_0^{1/2}[\theta(z-v_0t-d_0)-\theta(z-v_0t-d)]
\end{equation}
The results (\ref{T}-\ref{U}) are valid for $\tau \leq 1$ i.e. $\omega_pt
\leq \gamma_0^{1/2}$,or for the distancies along the plasma column $l \sim
ct \leq \frac{1}{2\pi}\lambda_p\gamma_0^{1/2}$.
From (\ref{U}) follows that in underdense plasma,when $E_0(\tilde{z}) 
\geq 0 $ inside the bunch,the bunch electrons are deaccelerated,except
the front electrons,which will always have the same velocity $v_0$,due to
$E_0(d)=0$.
In the overdense plasma and for bunches longer than plasma nonlinear wave 
length \cite{E}.
\begin{equation}
\label{V}
z_{\lambda}=\frac{2v_0}{\omega_p}\frac{(1+\alpha)^{3/2}(1+\alpha\beta_0)^
{1/2}}{1-\alpha\beta_0}E\left(\frac{\pi}{2},k\right)
\end{equation}
the electrons in the head part of the bunch,where $d-z_\lambda/2 \leq z 
\leq d$,
and $E_0(\tilde{z}) \geq 0$,are deaccelerated and in the rear part,where
$d_0 \leq \tilde{z} <d-z_\lambda/2$ and $E_0(\tilde{z})<0$ are accelerated.
(In (\ref{V}) 
$\alpha=\frac{n_b/n_0}{1-n_b/n_0},
k=\left(\frac{2\alpha\beta_0}{1+\alpha\beta_0}\right)^{1/2},
E\left(\frac{\pi}{2},k\right)$ is the second kind complete elliptic
integral).
Hence,the bunch traversing the overdense plasma contracted around the point
$\tilde{z}=d-z_\lambda/2$,and traversing the underdense plasma it always 
expands.The quantative estimates of these changes will be given bellow.
They are small in the considered ultrarelativistic bunch case.

The eqs. for the electron bunch momenta in the next approximations have 
the following form:
\begin{eqnarray}
\label{W}
\frac{\partial \rho_{b2}}{\partial \tau}+\beta_0\frac{\partial \rho_{b2}}
{\partial \zeta}=-E_1(\tilde{z},\zeta,\tau) \\ \nonumber
\frac{\partial \rho_{b3}}{\partial \tau}+\beta_0\frac{\partial \rho_{b3}}
{\partial \zeta}=-E_2(\tilde{z},\zeta,\tau) \\ \nonumber
\frac{\partial \rho_{b4}}{\partial \tau}+\beta_0\frac{\partial \rho_{b4}}
{\partial \zeta}=-E_3(\tilde{z},\zeta,\tau)
\end{eqnarray}
\begin{equation}
\label{X}
\frac{\partial \rho_{b5}}{\partial \tau}+\beta_0\frac{\partial \rho_{b5}}
{\partial \zeta}=-E_4(\tilde{z},\zeta,\tau)-\beta_{b5}\frac{\partial 
\rho_{b1}}{\partial \tilde{z}}
\end{equation}

The quantities $E_i(\tilde{z},\zeta,\tau),i=1,2,3,4$ entered in (\ref{W}),
(\ref{X}) must be found as a solutions of the subsequent system of the 
equations for plasma electrons,resulting from (\ref{B},\ref{D},\ref{F}).

\section{Continuity Equation for the Bunch Electrons}

The decompositions (\ref{J}),(\ref{L}) of the continuity eq (\ref{E}) gives
\begin{equation}
\label{Y}
\left(\frac{\partial}{\partial \tau}+\beta_0\frac{\partial}{\partial 
\zeta}\right)n_{bi}=0 \qquad i=1,2,3
\end{equation}

Initial conditions $n_{bi}=0$,when $\tau=0$,are satisfied only when 
$n_{bi}(\zeta-\beta_{b0}\tau)=0,i=1,2,3$.In the fifth order approximation,
eq. for $n_{b4}$ is 
\begin{equation}
\label{Z}
\left(\frac{\partial}{\partial \tau}+\beta_0\frac{\partial}{\partial \zeta}
\right)n_{b4}=-n_{b0}\frac{\partial \beta_{b5}}{\partial \tilde {z}}
\end{equation}
or using (\ref{P}),(\ref{U}),
\begin{equation}
\label{AA}
\left(\frac{\partial}{\partial \tau}+\beta_0\frac{\partial}{\partial \zeta}
\right)n_{b4}=\frac{n_{b0}}{\beta_0^3}\frac{\partial E_0}{\partial 
\tilde {z}}\tau[\theta(\zeta-\beta_0\tau-\epsilon d_0)-\theta(\zeta -
\beta_0\tau-\epsilon d)] 
\end{equation}
The solution of eq. (\ref{AA}),obeing the usual boundary and initial 
conditions,is
\begin{equation}
\label{AB}
n_{b4}=\frac{n_{b0}}{2\beta_{b0}^3}\frac{\partial E_0(\tilde{z})}{\partial 
\tilde {z}}\tau^2[\theta(\zeta-\beta_0\tau-\epsilon d_0)-\theta
(\zeta-\rho\tau-\epsilon d)]
\end{equation}
From (\ref{AB}) follows,that for the underdense plasma,when $\frac{\partial
E(z)}{\partial z}<0$ everywhere inside the bunch,$n_{b4}<0$ and the bunch 
expands.For the overdense plasma $n_{b4}$ changes sign:on the first quarter
of the plasma wave length $z_\lambda$ (\ref{V}) it is negative,then positiv 
on the
second and third quarters of the $z_\lambda$ and again negative on the forth 
quarter of the $z_\lambda$.Hence in the overdense plasma the bunch of the 
length $d \geq z_\lambda$ contracted on the middle of the bunch.Density
modulation with the period equal $z_\lambda$ for the long bunches $d > 
z_\lambda$ will take place. (See also \cite{H}).

Equations for the next approximations are
\begin{eqnarray}
\label{AC}
\left(\frac{\partial}{\partial \tau}+\beta_0\frac{\partial}{\partial \zeta}
\right)n_{b5}=-n_{b0}\left(\frac{\partial \beta_{b6}}{\partial \tilde{z}}+
\frac{\partial \beta_{b5}}{\partial \zeta}\right) \\ \nonumber
\left(\frac{\partial}{\partial \tau}+\beta_0\frac{\partial}{\partial \zeta}
\right)n_{b6}=-n_{b0}\left(\frac{\partial \beta_{b7}}{\partial \tilde{z}}+
\frac{\partial \beta_{b6}}{\partial \zeta}\right)
\end{eqnarray}
the solutions of the eqs. (\ref{AC}),using (\ref{P}-\ref{R}) with 
$\rho_{b2}=\rho_{b3}=0$ (see below),are
\begin{eqnarray}
\label{AD}
n_{b5}=\frac{n_{b0}}{\beta_0^3}\left[E_0(z)\frac{\partial E_0}{\partial
\tilde{z}}\tau^3+E_0(z)(\zeta-\epsilon d_0)\right] \times \\ \nonumber
[\theta(\zeta-\beta_0\tau-\epsilon d_0)-\theta(\zeta-\beta_0\tau-\epsilon d)]
\end{eqnarray}
\begin{eqnarray}
\label{AE}
n_{b6}=\frac{n_{b0}}{\beta_0^3}\left[-\frac{3}{2}\beta_0E_0^2(z)
\frac{\partial E_0}{\partial\tilde{z}}\tau^4+3E_0(\tilde{z})
(\zeta^2-2\epsilon\zeta d_0+\epsilon^2d_0^2)\right] \times \\ \nonumber
[\theta(\zeta-\beta_0\tau-\epsilon d_0)-\theta(\zeta-\beta_0\tau-\epsilon d)]
\end{eqnarray}
The second terms in the first square brackets in (\ref{AD}),(\ref{AE}) are
absent for the cases,when bunch is contracted,$d_0>0$.

The boundary condition $n_{bi}=0,p_{bi}=0$ at $\tilde{z}=d$ means that 
the front of the bunch always moves with the constant initial velocity
$v_0$ in the lab system (see also \cite{K}).The end of the bunch,which was
at $\tilde{z}=0$,when $t=0$,can be changed and be at $\tilde{z}=d_0 \neq 0$,
when $t>0$.From the bunch charge conservation it is possible to find
$d_0(t)$ using the relation:
\begin{equation}
\label{AF}
n_{b0}d=\int_{d0}^d{(n_{b0}+\epsilon^4n_{b4}+\epsilon^5n_{b5}+\epsilon^6n_{b6}+
\cdots)}d\tilde{z}
\end{equation}
An approximate expression for $d_0(t)$ follows from (\ref{AF}),using
(\ref{AB},\ref{AD},\ref{AE}):
\begin{eqnarray}
\label{AG}
d_0(t)=-\frac{t^2}{2\gamma_0^3}E_0(0)-\frac{t^3}{2\gamma_0^4}E_0^2(0)+
\frac{t^4}{2\gamma_0^5}E_0^3(0)+\\ \nonumber
+\frac{1}{\gamma^3}\int_0^d{E_0(\tilde{z})
(\tilde{z}+v_0t)}d\tilde{z}+\frac{3}{\gamma^4}\int_0^d{E_0^2(z)
(\tilde{z}^2+2\tilde{z}v_0t+(v_0t)^2)}d\tilde{z}
\end{eqnarray}
(in (\ref{AG}) all quantities are dimensionless,see (\ref{A})).
The last two terms in (\ref{AG}) are absent,when $d_0 < 0$.From
(\ref{AG}) it is evident that ultrarelativistic bunch practically does
not change it's length,passing through plasma during the time interval
$t \leq \omega_p^{-1}\gamma_0^{1/2}$,up to terms of the order of 
$\gamma_0^{-2}$.

\section{Dynamics of the Plasma Electrons}

The plasma electron motion,described by eq. (\ref{B}),(\ref{D}),(\ref{F}),
in the zero order approximation (fixed bunch,steady state regime) is 
considered in \cite{E}.In the next--first,second and third-- approximations,
due to the solutions of eq. (\ref{Y}) $n_{bi}(\zeta-\beta_{b0}\tau)=0,
i=1,2,3,$ it is necessary to choose the solutions of eq. (\ref{B}),
(\ref{D}),(\ref{F}),which are zero inside the bunch for all values of $\tau:$
$$n_{ei}=0,E_i=0,\rho_{ei}=0,i=1,2,3.$$
The nonzero solutions,which are possible to obtain analytically for small
$|\rho_{e0}| \ll 1$ and large $|\rho_{e0}| \gg 1$ values of plasma electron
momenta and which satisfy the zero boundary and initial conditions,are
physically meaningless.

System of the equations (\ref{B}),(\ref{D}),(\ref{F}) in the forth order
has the form
\begin{equation}
\label{AH}
\frac{\partial \rho_{e4}}{\partial \tilde{z}}-\frac{\rho_{e4}E_0(\tilde{z})
(1+\rho_{e0}^2)^{1/2}}{[\beta_0(1+\rho_{e0}^2)^{1/2}-\rho_{e0}]^2}=
E_4\frac{(1+\rho_{e0}^2)^{1/2}}{[\beta_0(1+\rho_{e0}^2)^{1/2}-\rho_{e0}]},
\end{equation}
\begin{equation}
\label{AI}
\frac{\partial E_4}{\partial \tilde{z}}=-n_{e4}-n_{b4}=-
\frac{n_0\beta_0(1+\rho_{e0}^2)^{1/2}\rho_{e4}}
{[\beta_0(1+\rho_{e0}^2)^{1/2}-\rho_{e0}]^2}-\frac{n_{b0}}{2\beta_0^3}
\frac{\partial E_0}{\partial \tilde{z}}\tau^2,
\end{equation}
where results of zero order approximation \cite{E},forth  order 
continuity eq. (\ref{D}),as well as relation $\beta_{e4}=\rho_{e4}
(1+\rho_{e0}^2)^{-1/2}$ are used.

Quasilinear system of eqs. (\ref{AH}),(\ref{AI}) can be treated numerically.
The analytical solution of the system (\ref{AH}-\ref{AI}) is possible
to obtain for small and large values of plasma electron momenta $\rho_e(0)$.
First consider the case of the small values of $|\rho_{e0}(\tilde{z})| 
\ll 1,\rho_{e0}(\tilde{z})<0$,which take place behind the front of the bunch
in the underdense and overdense plasma and around the $\tilde{z} \sim 
z_\lambda$ in overdense plasma.

In the considered case the eqs. (\ref {AH}-\ref{AI}) simplyfies due to 
relations,valid for $|\rho_{e0}| \ll 1,
\frac{\partial E_0(\tilde{z})}{\partial z} \approx -n_{0b},
E_0 \approx \pm\left[2\frac{n_{b0}}{n_0}\beta_0|\rho_{e0}|\right]^{1/2} 
\ll 1$
and the resulting eq. for $\rho_{e4}$ is:
\begin{equation}
\label{AJ}
\frac{\partial^2\rho_{e4}}{\partial \tilde{z}^2}+\frac{1}{\beta_0^2}
(n_{b0}+n_0)\rho_{e4}=\frac{1}{2\beta_0^3}n_{b0}^2\tau^2
\end{equation}
Solution of eq. (\ref{AJ}) with the boundary conditions
$$\rho_{e4}(\tilde{z}=d)=0,\frac{\partial \rho_{e4}(\tilde{z}=d)}
{\partial \tilde{z}}=0$$
which follows from $E_4(\tilde{z}=d)=0$ is the following:
\begin{equation}
\label{AK}
\rho_{e4}=\frac{n_{b0}^2\tau^2}{2\beta_0(n_0+n_{0b}})
\left[1-\cos\left(\frac{n_{b0}+n_0}{\beta_0^2}\right)^{1/2}
(d-\tilde{z})\right]
\end{equation}
Then from (\ref{AH}),(\ref{AI}):
\begin{equation}
\label{AL}
E_4 \approx \beta_0\frac{\partial \rho_{e4}}{\partial z}=
-\frac{n_{b0}^2\tau^2}{2\beta_0(n_0+n_{b0})^{1/2}}\sin\left(
\frac{n_0+n_{b0}}{\beta_0}\right)^{1/2}(d-\tilde{z})
\end{equation}
\begin{equation}
\label{AM}
n_{e4} \approx \frac{n_0}{\beta_0}\rho_{e4}=\frac{n_0n_{b0}^2\tau^2}
{2\beta_0^2(n_0+n_{b0})}\left[1-\cos\left(\frac{n_{b0}+n_0}{\beta_0}
\right)^{1/2}(d-\tilde{z})\right]
\end{equation}
It is essential to note that all corrections (\ref{AK},\ref{AL},\ref{AM})
to zero order approximations to the plasma electron momenta,density and
electric field inside the bunch are by the order of magnitude proportional
to $\epsilon^4\cdot\epsilon^2=\gamma_0^{-3}$ ($\epsilon^4$ is from forth
order,and $\epsilon^2$ from $\tau^2=\epsilon^2t^2$).Even for $\tau \sim 1$
corrections are by order of magnitude proportional to $\gamma_0^{-2}$.

Subsequent corrections to the wake wave behind the bunch,due to continuity
condition at the end of the bunch $(\tilde{z}=0)$,will be of the same order
of magnitude,i.e. the rigid bunch approximation for wake waves is valid 
up to terms,proportional to $\gamma_0^{-2} \ll 1$.

For the case $|\rho_{e0}| \gg 1$,which take place around $\tilde{z} \approx
\frac{z_\lambda}{2}$ for overdense plasma and at rear part of the long enough
bunch for underdense plasma,the approximate solutions of the forth order
problem are:
\begin{eqnarray}
\label{AN}
\rho_{e4}=-\frac{n_{b0}\tau^2}{2\beta_{b0}^2}\int_d^{\tilde{z}}E_0(\tilde{z})
d\tilde{z} \\ \nonumber
E_4=-\frac{n_{b0}\tau^2}{2\beta_{b0}^2}E_0(\tilde{z}) \\ \nonumber
n_{e4}=-\frac{n_0n_{b0}\tau^2}{2(1+\beta_{b0})^2\beta_{b0}|\rho_{e0}|}
\int_d^zE_0(\tilde{z})d\tilde{z}
\end{eqnarray}

For the underdense case $E_0(z) \sim |\rho_{e0}|^{1/2}$ and is large,so the
quantities in (\ref{AN}) could be large too for $\tau \leq 1$;but the 
Lorentz factor dependence remains the same as in the case of small
$|\rho_{e0}| \ll 1$.

Hence, only forth  order corrections to plasma electron momenta,electron
density and electric field inside and behind the bunch are differ from 
zero and are proportional to $\gamma_0^{-2}$,in the considered 
ultrarelativistic bunch case.

\section{Self-acceleration of the Bunch Electrons}

The back influense of the plasma wake waves on the bunch electron momenta
is calculated in the first approximation (see (\ref{T}-\ref{U})) in the
section 3. In the next approximations $\rho_{bi}=0,i=2,3,4$,due to $E_k=0,
k=1,2,3$ (see section 5) and the initial condition $\rho_{bi}(\tau=0)=0$.
From the sixth order equation it is possible to find from eqs. (\ref{T},
\ref{X},\ref{AM},\ref{AO}) $\rho_{b5}$ which differs from zero.For small
$|\rho_{e0}| \ll 1$ and large $|\rho_{e0}| \gg 1$ it is possible to calculate
$\rho_{b5}=b(\tilde{z})\tau^3$,where $b(\tilde{z})$ is a known function of
$\tilde{z}$.The subsequent contribution to bunch electron momenta is
$\epsilon^5\rho_{b5}=b\gamma_0^{-4}t^3$,which is much smaller in considered
case than first order contribution (\ref{U}) $\epsilon\rho_{b1}=
-E_0(z)\gamma_0^{-1}t$,even when $t$ approaches its limit value 
$t < \omega_p^{-1}\gamma_0^{1/2}$.

Consider the first order correction (\ref{T},\ref{U}) in more detail.
From (\ref{I},\ref{T},\ref{U}) in overdense plasma case the electrons
of the rear part of the bunch with the length $d > z_\lambda/2$ accelerate
and the increase of the momenta in the ordinary units is
\begin{equation}
\label{AO}
\triangle\rho_b=\rho_{b0}+\epsilon\rho_{b1}-\rho_{b0}=-eE_0(z)t
\end{equation}

The maximum of the electric field inside the bunch in overdense plasma
case is \cite{E}:
 $E_0^{max} \approx \frac{mc\omega_p}{e}$,
and
$$c\triangle\rho_b=mc^2\omega_pt,\frac{\triangle\rho_b}{\rho_{b0}}=
\frac{\omega_pt}{\gamma_0}$$ 
acceleration gradient 
$$G=\frac{c\triangle p}{el}=\frac{mc^2}{e}\frac{2\pi}{\lambda_p}=
\frac{\pi}{\lambda_p}Mv/cm,(l=ct).$$
If 
$$t \leq \omega_p^{-1}\gamma_0^{1/2}$$
then 
$$c\triangle p_b \leq 0,5\gamma_0^{1/2}Mev,\frac{\triangle p_b}{p_{b0}} 
\leq \gamma_0^{-1/2};$$
for 
$$cp_{b0}=10Mev,\gamma_0=20,c\triangle p_b \leq 2,23Mev,\frac{\triangle p_b}
{p_{b0}} \leq 22\%;$$
for 
$$cp_{b0}=100Mev,\gamma_0=200,c\triangle p_b \leq 7,07Mev,\frac{\triangle 
p_b} {p_{b0}} \leq 7\%.$$
For plasma densities 
$n_p=10^{13}\div10^{15}cm^{-3}$,accelerating gradient is $G \approx 
(1\div10) \frac{Mv}{cm}$.

It is possible to increase the value of selfaccelerated bunch electrons 
momenta using the combined bunch in the general form suggested
by M.L. Petrosian (privite communication).First part of the bunch has a 
density $n_b^{(1)} \gg \frac{1}{2}n_0$ (underdense case),and the density 
of the second part is $n_b^{(2)}<\frac{1}{2}n_0$ (overdense case).The first
part of the bunch has a length $d^{(1)},0 \leq \tilde{z} \leq d^{(1)}$ and
the second part $d^{(2)},-d^{(2)} \leq \tilde{z} \leq 0$.Then on the end 
of the first part of the bunch electric field in dimensionless variables 
is \cite{E}:
$$\frac{1}{2}E_0^2(0)=(n_b^{(1)}-n_0)\left[(1+\rho^2_{e0}(0))^{1/2}-1
\right]-n_b^{(1)}\beta_0\rho_{e0}(0)$$
When $\rho_{e0}(\tilde{z}=d)=0,E_0(\tilde{z}=d)=0$ and increases,when 
$|\rho_{e0}|$ increases $(\rho_{e0}<0)$.The largest value of $|\rho_{e0}|$
can be estimated from total momenta conservation,which has the form:
\begin{eqnarray}
\label{AP}
 n_b^{(1)}\rho_{e0}d^{(1)}\gamma_0 \geq \int_0^{d_1}\rho_{e0}(\tilde{z})
n_{e0}(\tilde{z})d\tilde{z} \approx \\ \nonumber
\approx \frac{n_0\beta_0}{1+\beta_0}\cdot\frac{|\rho_{e0}^{max}|d^{(1)}}{2}
\approx \frac{n_0d^{(1)}|\rho_{e0}^{max}|}{4}
\end{eqnarray}
It was taken into account,that \cite{E}
\begin{equation}
\label{AQ}
n_{e0}(\tilde{z})=\frac{n_0\beta_0(1+\rho_{e0}^2)^{1/2}}
{\beta_0(1+\rho_{e0}^2)^{1/2}-\rho_{e0}} \rightarrow \frac{n_0\beta_0}
{1+\beta_0}
\end{equation}
when $|\rho_{e0} \gg 1$;$(n_e\rho_{e0})$ dependence on $\tilde{z}$ is 
approximated as linear one.From (\ref{AP}) follows that 
\begin{equation}
\label{AR}
|\rho_{e0}^{max}| \leq \frac{4n_b^{(1)}\rho_{b0}\gamma_0}{n_0}
\end{equation}

The electric field in the second part of the combined bunch,where $n_b^{(2)}<
\frac{1}{2}n_0$ is given by the expression:
\begin{eqnarray}
\label{AS}
E_0(\tilde{z})=\pm\left\{E_0^2(0)+2\left(n_0-n_b^{(2)}\right)\right. 
\times \\ \nonumber \times \left. \left[\left(
1+\rho_{e0}^2(0)\right)^{1/2}-\left(1+\rho_{e0}^2\right)^{1/2}\right]-
2n_b^{(2)}\beta_{b0}(\rho_{e0}-\rho_{e0}(0))\right\}^{1/2}= \\ \nonumber
=\pm\left[2\left(n_0-n_b^{(2)}\right)\left(a-\left(1+\rho_{e0}^2\right)^
{1/2}-\alpha^{(2)}\beta_0\rho_{e0}\right)\right]^{1/2},
\end{eqnarray}
where
$$\alpha^{(2)}\equiv\frac{n_b^{(2)}}{n_0-n_b^{(2)}},$$
$$a\equiv 1+\frac{n_b^{(1)}-n_b^{(2)}}{n_0-n_b^{(2)}}\left[\left(1+\rho_{e0}^2
(0)\right)^{1/2}-1\right]-\frac{n_b^{(1)}-n_b^{(2)}}{n_0-n_b^{(2)}}
\beta_0\rho_{e0}(0);$$
$a \rightarrow 1$,when $\rho_e(0) \rightarrow 0,a > 1$,when $\rho_e(0) < 0$,
and
\begin{equation}
\label{AT}
 a \approx \frac{n_b^{(1)}-n_b^{(2)}}{n_0-n_b^{(2)}}(1+\beta_0)|\rho_{e0}|
\gg 1,
\end{equation}
when $|\rho_{e0}(0)| \gg 1$.

Electric field $E_0(\tilde{z})$ (\ref{AS}) is equal zero at
\begin{equation}
\label{AU}
\rho_{\pm}=-\frac{a\alpha^{(2)}\beta_0}{1-(\alpha^{(2)}\beta_0)^2}
\pm \left[\left(\frac{a\alpha^{(2)}\beta_0}{1-(\alpha^{(2)}\beta_0)^2}
\right)^2+
\frac{a^2-1}{1-(\alpha^{(2)}\beta_0)^2}\right]^{1/2}
\end{equation}
and when $|\rho_{e0}(\tilde{z})| > |\rho_{-}|$ ($\rho_{-}$ is the root of 
(\ref{AR})
with minus sign in (\ref{AU})) the bunch electron selfacceleration can
take place in the region of the bunch,where $E_0(\tilde{z}) < 
0.(E_0(\tilde{z}) <0,$
when $\rho_{e0}(\tilde{z})$ decreases with increasing $\tilde{z}$ \cite{E}. 
Field $|E_0(\tilde{z})|$ (\ref{AR}) has a  maximum value,when
\begin{equation}
\label{AV}
\rho_e=\rho_e^{max}=-\frac{\alpha^{(2)}\beta_0}{\left[1-(\alpha^{(2)}\beta_0)^
2\right]^{1/2}}
\end{equation}
For large value of $|\rho_e(0)| \gg 1,a \gg 1$ from (\ref{AT})
$\rho_{-} \approx -\frac{a}{1-\alpha^{(2)} \beta_0}$
and conditions $|\rho_{-}| > |\rho_e(0)|,|\rho_{-}| > |\rho_e^{max}|$ are 
always
fulfilled.As a cons
equence,$E_0(\tilde{z})$ in the second part of the 
combined bunch decreases from $E(0) > 0$ to zero,then became negative,reaches
its minimum value (it modulus is maximum) and then tends to zero at 
$\tilde{z}=
-\tilde{z}_\lambda^{(2)}$.Maximum value of the modulus of the electric field
is
\begin{equation}
\label{AW}
E_0^{max} \approx \left[2(n_0-n_b^{(2)})a\right]^{1/2} \approx
\left[4(n_b^{(1)}-n_b^{(2)})|\rho_e(0)|\right]^{1/2}
\end{equation}

Unfortunately,the plasma electron momenta largest value (\ref{AR}) is
overestimated.More restrict estimate follows from the condition 
$n_e(\tilde{z}) \geq 0$ (\ref{AQ}).It is so for any $\rho_{e0} < 0$,
which is the case everywhere inside the bunch,but in wake field $\rho_{e0}$
can be positiv too,and from $n_{e0}(\tilde{z}) > 0$ follows that 
$0 \leq \rho_{e0}(\tilde(z) \leq \beta \gamma_0$ instead of condition
(\ref{AR}).If $ \rho_{e0}(\tilde{z}) > \beta \gamma_0,n_e$ became negative,
which means that basic assuption on the existence of the steady state
regime is violated and this case needs the development of a new approach,
which is nonstationary from the beginning.

It is possible to see that in wake field
\begin{equation}
\label{AX}
-\left[\left(\frac{1}{n_0}A\right)^2-1\right]^{-1/2} \leq 
\rho_{e0}(\tilde{z}) \leq \left[\left(\frac{1}{n_0}A\right)^2-1\right]^{1/2}
\end{equation}
and subsequently
\begin{equation}
\label{AY}
1 \leq \left(\frac{1}{n_0}A_0\right) \leq \gamma_0,
A=(n-n_b^{(2)})a+n_b^{(2)}[1+\rho_{e0}^2(-d_2)]^{1/2}-n_b^{(2)}\beta_0
\rho_{e0}(-d_2),
\end{equation}
where $"a"$ is given by (\ref{AS}) and (\ref{AT}) for large value of 
$|\rho_{e0}(0)| \gg 1$.If $$|\rho_{e0}(-d_2)| <1, n_b^{(1)} \gg n_b^{(2)}$$
\begin{equation}
\label{AZ}
\frac{1}{n_0}A \approx 2\frac{n_b^{(1)}}{n_0}|\rho_{e0}(0)| \leq \gamma_0
\end{equation}
and
\begin{equation}
\label{BA}
E_0^{max}=[4(n_b^{(1)}-n_b^{(2)})|\rho_{e0}|]^{1/2} \leq (2n_0\gamma_0)^{1/2}
\end{equation}

The increase of the bunch electron momenta and acceleration gradient in the
ordinary units then are
\begin{eqnarray}
\label{BB}
\triangle p_b \leq (2p_{b0}mc)^{1/2}\omega_p t < 2^{1/2}mc\gamma_0 \\ 
\nonumber
\frac{\triangle p_b}{p_b} \leq \left(\frac{2}{\gamma_0}\right)^{1/2}\omega_pt
< (2)^{1/2}
\end{eqnarray}
\begin{equation}
\label{BC}
G=\frac{c\triangle p_b}{el} \leq \frac{2\pi}{\lambda_p}\frac{(2cp_{b0}mc^2)^
{1/2}}{e}=\frac{2\pi}{\lambda_p}{mc^2(2\gamma_0)^{1/2}}{e}^{-1},
l=ct.
\end{equation}
In (\ref{BB}),(\ref{BC}), it is taken into account that these relations are 
valid for
$\omega_pt \leq \gamma_0^{1/2}$ and subsequent length of the plasma 
column must be $l \sim ct \leq \frac{\lambda_p}{2\pi}\gamma_0^{1/2}$.
So the total increase of the momenta is not exited the initial value of 
that,but acceleration gradient increases proportionally to $\gamma_0^{1/2}$
and for example for $n_0=3\cdot10^{13}\div 3\cdot10^{15}cm^{-3}$ and for
$\gamma_0=20,0 
({\cal{E}}_0=10,0Mev);\gamma_0=200,0;({\cal{E}}_0=100,0Mev); 
\gamma_0=2000,0 ({\cal{E}}_0=2Gev),G$ is $G=(31,5\div 315)\frac{Mv}{cm};
(100\div 1000)\frac{Mv}{cm};(223 \div 2230)\frac{Mv}{cm};$ subsequently.

Selfacceleration of the electrons from the second part of the combined
bunch is accompanied by deacceleration at the same extend the electrons
from the rear of the first part of the combined bunch.In the both cases
the change of the velocities according to (\ref{O},\ref{P},\ref{T},
\ref{BA}) is negligible for $\gamma_0 \gg 1$
$$\beta_b=\beta_0 \pm \gamma_0^{-2}(2n_0)^{1/2}$$

Hence,the combined bunch will not changes essentially its shape and density
distribution during the time of selfacceleration.The estimate (\ref{BA})
of the maximum electric field on the end of the first part of combined
bunch is true for long enough length $d_1$.Using results,obtained in 
\cite{O} for $|\rho_{e0}| \gg 1$ and $1 \ll \frac{n_b^{(1)}}{n_0} \ll 
2\gamma_0^2$,it is possible to estimate the length $d_1$,which in ordinary
units is
\begin{equation}
\label{BD}
d_1 \leq \frac{\lambda_p}{2\pi}\frac{n_0}{n_b^{(1)}}\gamma_0^{1/2}
\end{equation}

For $$\gamma_0=20,200,2000;n_0=3\cdot10^{11};n_b=3\cdot10^{12}cm^{-3};$$
$$d_1 \leq 0,07\lambda_p=0,44cm;0,22\lambda_p=1,38cm;0,7\lambda_p=4,4cm$$
subsequently.

It is useful to mention,that the considered mechanism of selfacceleration
has much in common with the selfacceleration of electrons in passive
resonant systems (resonant cavities,wave guides) which was widely 
investigated
both theoretically and experimentally in the seventies (see e.g. \cite{P},
\cite{Q} and review in \cite{R};review of more recent results is presented
in \cite{S}).

\section{Conclusion}

The multiple scale perturbative approach \cite{M} is applied to the one-
dimensional problem of the buck influence of the plasma wake field on 
bunch electrons and wakes itself for the ultrarelativistic bunch.Inverse 
square root of the bunch electrons Lorentz factor is choosen as a small 
parameter $\epsilon=\gamma_0^{-1/2}$.It is shown that plasma wake fields 
characteristics experinced a perturbations only in the forth order terms,
proportional to $\epsilon^4=\gamma_0^{-2}$,so are small for considered 
ultrarelativistic case.It means that rigid bunch approximation is valid
for ultrarelativistic bunch up to the terms,proportional to $\gamma_0^{-2}$.

The momenta of the bunch electrons changes in the first order approximation,
eqs. (\ref{T},\ref{U}),density and bunch length in the forth order,eqs.
(\ref{AB},\ref{AF},\ref{AG}).The bunch traversing the overdense plasma
contracted around the point $\tilde{z}=d-z_\lambda/2$, and expands when 
traverses underdense plasma.

The change in the bunch electrons momenta
is more essential.In the cases of underdense plasma $(n_b < \frac{1}{2}n_0)$
or in the model example considered a combined bunch (first part of the
bunch with the uniform density $n_b^{(1)} \gg \frac{1}{2}n_0$,the second
part with the density $n_b^{(2)} \gg \frac{1}{2}n_0$) the remarkable
selfacceleration of the bunch electrons can take place.

Predicted selfacceleration can be tested experimentally and it may be of
the practical importance.It is worthwhile to mention,that Langmuir already
noticed,that the beam,which has passed the plasma column,contains a
significant portion of electrons with energies higher than the initial
energy.

In the recent times various groups (see e.g. \cite{T},\cite{U},\cite{V})
observed the effect of selfacceleration experimentally.

It is possible to hope,that the present work will stimulate more systematic
study of the selfacceleration of the electrons of the bunches,moving in
plasma.

As it follows from the presented consideration the different approaches
to the problem out of the presented frame of steady state regime 
taken as a first order approximation,may offer a new possibility to 
selfaccelerate the electrons of the bunches,moving in plasma.

\section{Aknowledgement}

Author is indebted to M.L. Petrosian,S.G. Arutunian,S.S. Elbakian,
A.G. Khachatrian,E.V. Sekhpossian for helpful discussions,suggestions and
comments.

\end{document}